\documentclass[smallextended,twoside]{article}   
\usepackage{draftwatermark}
\usepackage[margin=2cm]{geometry}
\pdfoutput=1 
\usepackage[hidelinks]{hyperref}
\usepackage{graphicx, amsmath, amssymb, amsfonts}
\usepackage[round]{natbib}
\usepackage{authblk}
\usepackage{fancyhdr}
\usepackage{algorithm2e}
\usepackage{datetime}
\newcommand{\dgvel}{\ensuremath{\mathrm{P1}_{DG}}}

\newdateformat{mydateformat}{%
\ordinaldate{\THEDAY}\ \monthname[\THEMONTH] \THEYEAR}

\SetWatermarkText{PRE-PRINT}
\SetWatermarkScale{4}

\begin{document}
\pagestyle{fancy}
\fancyhead{}
\fancyhf[cf]{}
\fancyhead[LE,RO]{\small{\thepage}}
\fancyhead[RE]{\small{Creech}}
\fancyhead[LO]{\small{Efficient Large Eddy Simulation for the Discontinuous Galerkin Method on unstructured meshes}}
\fancyfoot{}

\title{\textbf{Efficient Large Eddy Simulation for the Discontinuous Galerkin Method}}

\author[1]{Angus Creech%
	\thanks{Corresponding author. Email: a.creech@ed.ac.uk }}
\author[2]{Adrian Jackson}
\author[3]{James Maddison}
\author[4]{James Percival}
\author[1]{Tom Bruce}

\affil[1]{Institute of Energy Systems, University of Edinburgh, Scotland}
\affil[2]{EPCC, University of Edinburgh}
\affil[3]{School of Mathematics and Maxwell Institute for Mathematical Sciences, University of Edinburgh}
\affil[4]{Dept. of Earth Science and Engineering, Imperial College London}

\date{\mydateformat\today}

\maketitle
\thispagestyle{empty}

\section*{Abstract}

In this paper we present a new technique for efficiently implementing Large Eddy Simulation with the Discontinuous Galerkin method on unstructured meshes. In particular, we will focus upon the approach to overcome the computational complexity that the additional degrees of freedom in Discontinuous Galerkin methods entail. The turbulence algorithms have been implemented within Fluidity, an open-source computational fluid dynamics solver. The model is tested with the well known backward-facing step problem, and is shown to concur with published results.


\maketitle

\section{Introduction}

The finite element method (FEM) is naturally well-adapted to problems with complex geometry, through the use of unstructured meshes. Two common formulations used in computational fluid dynamics are Continuous Galerkin (CG) and Discontinuous Galerkin (DG). CG finite element methods typically approximate numerical solutions with functions which are continuous and differentiable almost everywhere, but with discontinuous derivatives permitted at the boundaries of mesh elements. One class of CG finite element function space is the space of P$n$ functions, which are polynomials of degree $n$ within mesh elements, and continuous at element boundaries. DG finite element methods, on the other hand, break this continuity, by having interpolation functions that are continuous only within the element, and discontinuous at element boundaries. One class of DG finite element function space is the space of P$n_{DG}$ functions, with polynomials of degree $n$ within mesh elements, which are permitted to be discontinuous at element boundaries.

DG finite element methods are related to the finite volume method (FVM), as local conservation properties inherent in the FVM discretisation of flux-divergence equations can also be guaranteed in P$n_{DG}$ discretisations. However, higher order FVM discretisations typically require ever larger stencils, using information from ever wider patches of mesh elements, which leads to a considerable increase in computational complexity. By contrast, high order DG discretisations can be constructed using basis functions which are only non-zero within a single mesh element, the higher order accuracy achieved through higher order polynomials. In this sense, higher order DG discretisations can be formulated while retaining spatial locality.

When discussing finite element discretisations of the Navier-Stokes equations it is common to discuss a finite element pair, which defines the types of functions which are used to approximate velocity and pressure. We consider numerical simulations of the incompressible Navier-Stokes equations using the \dgvel-P2 finite element pair \citep{cotter2009,cotter2011,cotter2009a}, which uses a discrete velocity whose components are linear within mesh elements but possibly discontinuous at element boundaries, and a discrete pressure which is quadratic within mesh elements and continuous at element boundaries. The choice of finite element pair is limited by the well-known condition for Ladyzhenskaya-Babu{\v{s}}ka-Brezzi stability (LBB) stability \citep{ladyzhenskaya1969,babuska1973,brezzi1974}, which implies that the discrete divergence of any non-trivial discrete pressure gradient must be appropriately bounded away from zero as the mesh is refined. The \dgvel-P2 finite element pair is LBB stable \citep{cotter2009}.

This paper details the application of the high-fidelity turbulence modelling technique  Large Eddy Simulation (LES) \citep{smag1963,deardroff1970} to Discontinuous Galerkin elements. The calculation of eddy viscosity in LES can pose a problem as DG spaces are locally but not globally differentiable. Moreover, the additional degrees of freedom in DG increase the computational complexity when compared to CG. Below we discuss how these problems can be approached to produce an efficient and stable turbulence modelling algorithm.

\section{Model methodology}
\label{s:Method}

\subsection{Reason d'etre}
\label{s:reasoning}

Modern computing hardware is reliant on fast, processor local memory, called caches, to realise good compute performance.  This reliance arises from the disparity in performance between processors and main memory, with processors capable of executing 100s to 1000s of instructions in the time it takes to load some data from main memory.

Caches exploit the principles of {\it temporal} and {\it spatial} data locality that most programs exhibit.   That is to say, most programs will either use the same piece of data multiple times for computations within a given time frame (temporal locality), or they will use data located close (in memory) to data that has been recently used.  Storing recently used data in the cache, and loading a larger region of data into cache when a specific item is required to complete an instruction, rather than simply loading that piece of data, enables processors to achieve high performance for a wide range of programs whilst still operating with a large and slow main memory system.

This being the case, it is essential for good computational performance of an implementation of an algorithm to be structured in such a way as to make good use of processor local cache memory.  There are a number of design techniques that can be used to increase efficient use of caches by programs; examples such as cache blocking, strip mining, and pre-fetching are widely exploited by applications and programs. 

Whilst it is the case that one of the DG method's strengths is the local nature of the memory access for element operations, its discontinuous nature becomes a problem for LES. In LES, a smoothing filter must be applied for numerical stability; this necessitates accessing and changing values held in neighbouring elements. By contrast, the CG method automatically has a filter-type effect by the virtue that field values at element edges are shared. Moreover, assuming mesh is relatively well ordered, nodal values in elements are more likely to already be in the prior to calculation, if they have previously been accessed by their neighbours. It is these twin virtues we aim to exploit in this hybrid DG-CG LES formulation, without sacrificing the numerical stability of DG methods when simulating turbulent flow.

\subsection{Basic equations}
\label{s:basic-equations}

We start with the Navier-Stokes momentum and continuity equations for incompressible Newtonian fluids in matrix form, ie.

\begin{equation}
\label{eqn:ns-eqn}
\frac{D u_i}{D t} = - \frac{1}{\rho} \frac{\partial p}{\partial x_i} 
    + \frac{\partial}{\partial x_j} \left[ \nu \left( \frac{\partial u_i}{\partial x_j} + \frac{\partial u_j}{\partial x_i} \right) \right] 
\end{equation}

\begin{equation}
\label{eqn:cont}
\frac{\partial u_i}{\partial x_i} = 0
\end{equation}

where $u_i$ is the velocity field, $\rho$ is the constant fluid density and $p$ is pressure. When this is discretised, it becomes necessary to parameterise the subgrid turbulence. In Large Eddy Simulations (LES), this is achieved through the addition of a sub-grid viscosity, $\nu_{sgs}$, which represents turbulence occurring at smaller, unresolved scales, so that (\ref{eqn:ns-eqn}) becomes

\begin{equation}
\frac{D \tilde{u}_i}{D t} = - \frac{1}{\rho} \frac{\partial p}{\partial x_i} 
    + \frac{\partial}{\partial x_j} \left[ \left( \nu + \nu_{sgs} \right) \left( \frac{\partial \tilde{u}_i}{\partial x_j} + \frac{\partial \tilde{u}_j}{\partial x_i} \right) \right] 
\end{equation}

where $\tilde{u}_i$ denotes the resolved velocity field. The next section will deal with the formulation of $\nu_{sgs}$ within a Discontinuous Galerkin context.

\subsection{Calculation of sub-grid viscosity}

Consider a mesh consisting of a set of elements $E$. Let the velocity components $u_i$ each be $\mathrm{P}1_{DG}$ functions: linear within the elements of this mesh $e \in E$, but possibly discontinuous at the element boundaries.

We define the sub-grid viscosity in its Smagorinsky-Lilly form \citep{smag1963, deardroff1970}, combined with a wall damping function, ie.

\begin{equation}
\label{eqn:sgs_orig}
  \nu_{sgs} = \left( C_s L \right)^2  D \|S\|
\end{equation}
$C_s$ is the Smagorinksy coefficient, usually set to $0.1$, and $L$ is the length scale of the subgrid scale filter, typically twice the gridsize. $D$ is a Van Driest-type wall damping factor in a similar vein to that of \citet{balaras1994, cabot1996,cabot1999}, which ensures that the eddy viscosity necessarily disappears at a no-slip boundary:

\begin{equation}
D=\left[1-e^{-y^+ / A}\right]^2
\end{equation}

with $y^+ $ as the distance from the no-slip boundary in wall units, and $A=25$. The last part of equation (\ref{eqn:cont}) is $\|S\|$, which we define as the 2-norm of the rate-of-strain tensor, ie.
\begin{equation}
  \|S\|  = \frac{1}{2} \| \left(J+J^T\right) \|
\end{equation}

where $J$ is the Jacobian of the velocity field. This presents several problems. Firstly, this Jacobian $J$ cannot be defined directly in terms of derivatives of the \dgvel\ velocity components as they are not in general appropriately differentiable. Secondly, a DG formulation of $J$ would imply a DG $\nu_{sgs}$: this adds considerable computational complexity, exacerbated by the need for a smoothing filter to be applied to $\nu_{sgs}$ to improve stability.

\begin{figure}
\centering
  \includegraphics[width=0.30\linewidth]{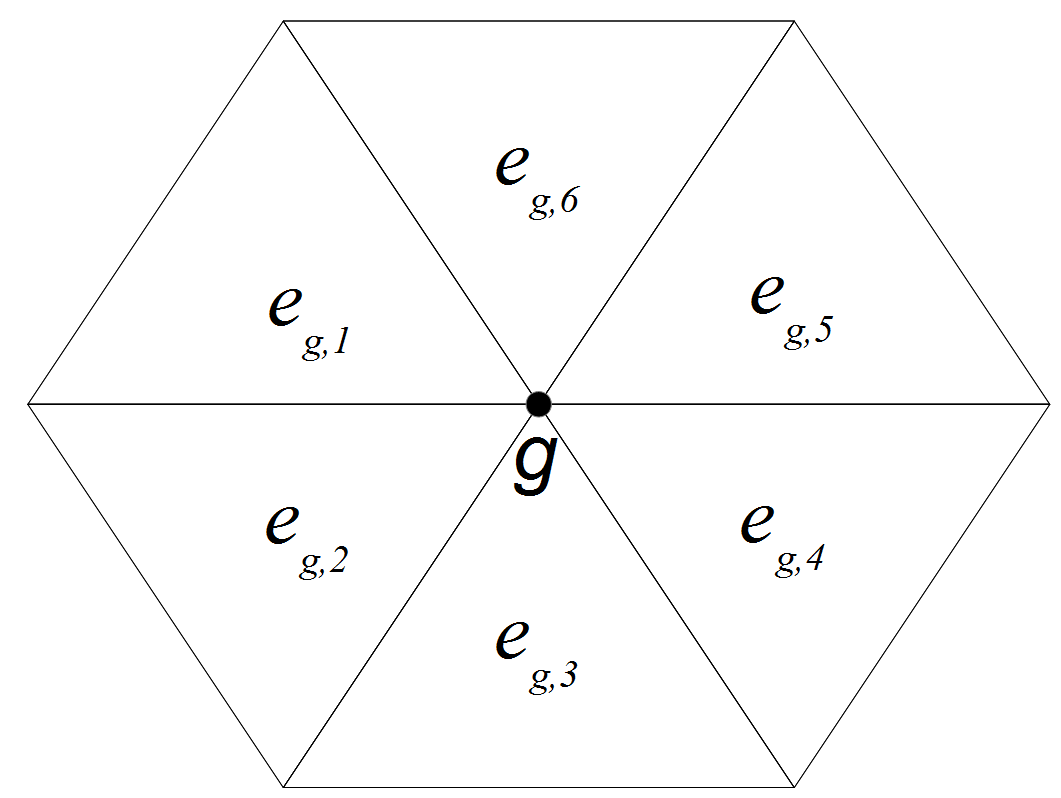}
	\caption{A two dimensional triangular CG mesh showing elements $e_{g,k}$ sharing global node $g$.}
	\label{fig:node-sharing-elements}
\end{figure}

These problems are here resolved by first projecting the velocity components to a P1 function space, which we will call $V$. Both the original DG space $U$ and the new CG space $V$ share the same mesh, with elements $E$, such that any mesh element $e \in E$. The projection is performed via a Galerkin $L^2$ projection \citep[][appendix A.3.2.1]{gresho2000}, which is a relatively cheap operation within Fluidity \citep{fluidity2005}, typically taking about 0.3\% of execution time per timestep. This allows the definition of a per-element sub-grid viscosity $\hat{\nu}_{sgs}$ which is piecewise constant within the mesh elements. Dropping the $sgs$ subscript for convenience, for element $e$ this is equal to
\begin{equation}
  \overline{\nu}_{pe}(e) = \left( C_s L \right)^2 \|S(e)\|
\end{equation}
where $L=L(e)$ is the lengthscale of the filter for element $e$, set to $L(e) = 2\Omega^{1/3}$, and $\Omega=\Omega(e)$ is the volume of the element. The rate-of-strain magnitude for this element is
\begin{equation}
  \|S(e)\|  = \frac{1}{2} \| \left(J(e)+J(e)^T\right) \|
\end{equation}
where $J(e)$ is the Jacobian of the velocity in element $e$, defined via the derivatives of the P1 projected velocity components in $V$. Note that as we are dealing with the derivatives of a P1 velocity, this is equivalent to the average value of the rate-of-strain across the element.

This, however, leaves us with a piecewise continuous subgrid viscosity. It is convenient to map this to a continuous P1 space, to yield a viscosity which is linear within the mesh elements and continuous at element boundaries. Let $N$ be a matrix defined so that $N(g,e)$ is equal to one if mesh node $g$ is connected to mesh element $e$, and zero otherwise. Without wall damping, the subgrid viscosity is defined at the mesh nodes as

\begin{equation}
\nu^{*}_{sgs}(g) = \alpha \left[ \frac{\sum_e N(g,e)\,\overline{\nu}_{pe}(e)}{\sum_e N(g,e)} \right] + (1-\alpha) \left[\frac{\sum_e N(g,e)\,\Omega(e) \overline{\nu}_{pe}(e)}{\sum_e N(g,e)\,\Omega(e) } \right]
\end{equation}

This first term is a local average of the viscosity values attached to node $g$ shown in Figure \ref{fig:node-sharing-elements}; this simply maps the piecewise continuous $\overline{\nu}_{ele}$ values to a continuous P1 field. The second term is a volume-weighted average of the viscosity values attached to node $g$, so that larger neighbouring elements exert greater influence on shared nodes, and is equal to the nodal value of a lumped mass Galerkin $L^2$ projection of $\overline{\nu}_{pe}$ (see \citet{gresho2000a} section 4.2.1). It can be considered as the filter part, applying an additional smoothing to the final nodal sub-grid viscosity, which is controlled via a pre-defined constant $\alpha$, where $0 \leq \alpha \leq 1$. For this paper, we set $\alpha=1/2$. 

Lastly, we add the wall damping term to give the final subgrid viscosity:

\begin{equation}
\nu_{sgs}(g) = D(g)\, \nu^{*}_{sgs}(g)
\end{equation}

where $D(g)=\left[1-e^{-y^+(g) / A}\right]^2$. 

\section{Test case: backward facing step}

\begin{figure}
\centering
  \includegraphics[width=0.85\linewidth]{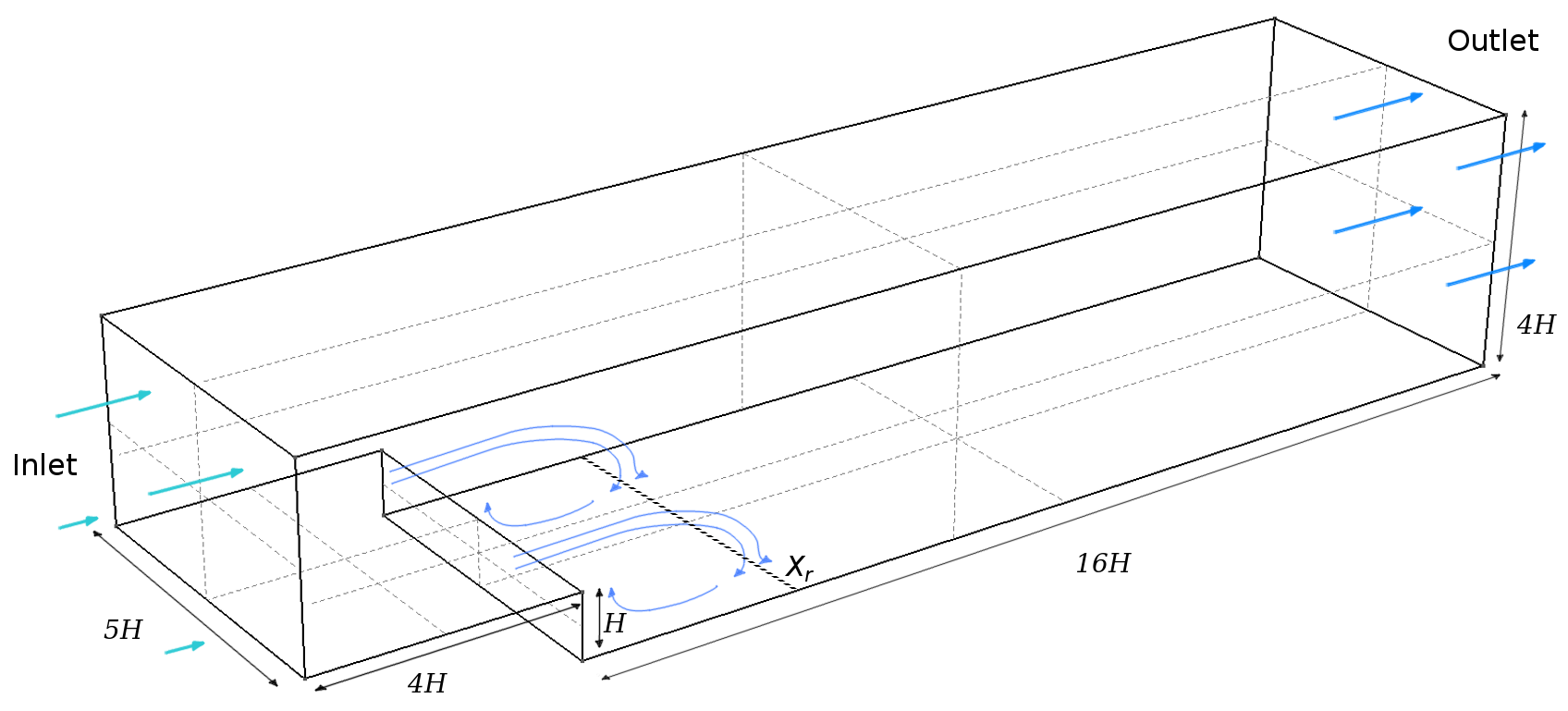}
	\caption{Backward facing step model with dimensions, showing the reattachment point $X_r$ beyond the step, the dividing line between recirculation and downstream flow.}
	\label{fig:backward-facing-step}
\end{figure}

To test the LES formulation, the well-known backward facing step was used, shown in Figure \ref{fig:backward-facing-step}. The backward facing step used here consists of a rectilinear channel, with a step of height $H$ and length $4H$ at the inlet. The channel is $20H$ long, $4H$ high, and $5H$ high. Following from \cite{jovic1994, aider2006}, we set the Reynolds number defined by the step height $H$, as $\mathrm{Re}_H=5000$. The domain was represented by an unstructured mesh of 913 060 tetrahedral cells, with the resolution increasing considerably near the no-slip bottom boundary. Elements far away from the no-slip boundary become mildly anisotropic, with a maximum horizontal/vertical aspect ratio of 4:1, which is within the limits for isotropic LES turbulence models \citep{scotti1993}.

With the \dgvel-P2 element pair previously discussed, the velocity elements used the Compact Discontinuous Galerkin formulation \citep{peraire2008}. A twice-Picard iterated approximation to the Crank-Nicolson scheme \citep{ford2004i} was used for time-stepping, together with adaptive time-stepping and advective subcycling. Stability tests gave a maximum allowable Courant number for timesteps of 1.9, with the subcycling Courant number set to 1.0.

Regarding boundary conditions, the bottom of the channel has a no-slip condition, and the sides and top are frictionless (free-slip). Similar to \cite{patil2012}, the inlet used the synthetic eddy method \citep{jarrin2006} to impose turbulent fluctuations on top of a logarithmic velocity profile, whilst the outlet is an open boundary.

In the synthetic eddy method, the turbulent inflow conditions used a log law for the mean velocity profile, which was based upon channel flow measurements by \citet{nezu1993}:

\begin{equation}
\overline{u}(z_s)=\left(\frac{u_\tau}{K}\right) \log\left(\frac{z_{ref}+z_R}{z_R}\right) + 8.5
\end{equation}

where $u_\tau$ is the frictional velocity, $K\approx 0.41$ is the von Karman constant, $z_s$ is the height above the step, $z_{ref}$ is the height at which we reference the free-stream flow speed $u_0$, and $z_R$ is the roughness height. The mean cross-stream and vertical components were set so that $\overline{v}=\overline{w}=0$.

We can define $u_\tau$ as

\begin{equation}
u_\tau =\left(\frac{u_{ref}}{K}\right) \log\left(\frac{z_{ref}+z_R}{z_R} + 8.5\right)^{-1}
\end{equation}

For our simulations, the roughness height was chosen to be very small compared to the step, with $z_R=0.0001$. The reference height for $u_0$ was chosen to be $z_{ref}=H/4$. The diagonal components of the Reynolds Stress $Re_{ii}$ need to be specified for the synthetic eddy methods; these were based on \citet{stacey1999}, which follows on from \citet{nezu1993}, so that

\begin{equation}
R_{uu} = \overline{u'u'} = 5.28 \, u^2_\tau \exp\left(-\frac{2z}{H_{in}}\right)
\end{equation}
\begin{equation}
R_{vv} = \overline{v'v'} = 2.66 \, u^2_\tau \exp\left(-\frac{2z}{H_{in}}\right)
\end{equation}
\begin{equation}
R_{ww} = \overline{w'w'} = 1.61 \, u^2_\tau \exp\left(-\frac{2z}{H_{in}}\right)
\end{equation}

where $H_{in}$ is the height of the channel at the inlet. The lengthscale components, $L_{u}$, $L_{v}$ and $L_{w}$ were taken from \citet{nezu1993}, and so for the bottom half of the inlet, the streamwise eddy length scale is

\begin{equation}
L_u = \sqrt{zH_{in}}
\end{equation}
With $L_u=\frac{1}{2}H_{in}$ for the top half. The cross-stream and vertical components are taken as $L_v=\frac{1}{2}\,L_u$ and $L_w=\frac{1}{4}\,L_u$ respectively.
\section{Results and discussion}
\label{s:results-discussion}

\begin{figure}
\centering
  \includegraphics[width=0.5\linewidth]{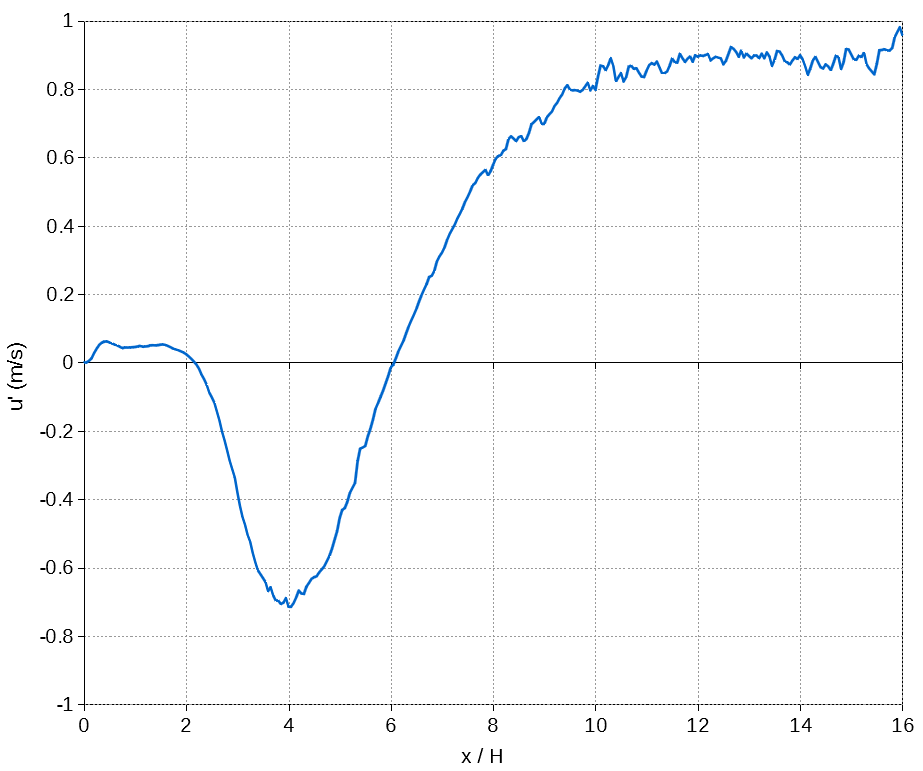}
	\caption{The normalised profile of time-averaged x-component of velocity, $u'$, plotted against the scaled distance downstream from the backstep, $x/H$. The profile sample points are in line downstream from the backstep to the outline, near to the no-slip boundary at the bottom of the domain. The mean recirculation point is clearly shown at $X_r=6.065 H$.}
	\label{fig:results-recirculation-point}
\end{figure}

The entire simulation domain including velocity, pressure and subgrid eddy viscosity, was saved every 60 simulated seconds. To ensure that results were statistically steady, the simulation finish time was set to 2.77 days. The first 200 dumps of simulation data were discarded; this represented the minimum spin-up time that demonstrated no effect on statistical calculations. To ascertain the recirculation point, $X_r$, measurements were taken in line from the cross-stream centre, running downstream from backstep to the outlet, just above the bottom of the domain. Four hundred samples were taken at equidistant points along this line, and the values for the x-component of the velocity, $u$, time-averaged to give mean profiles over the sampling period: the normalised results are shown in Figure \ref{fig:results-recirculation-point}. 

\begin{figure}
\centering
  \includegraphics[width=0.99\linewidth]{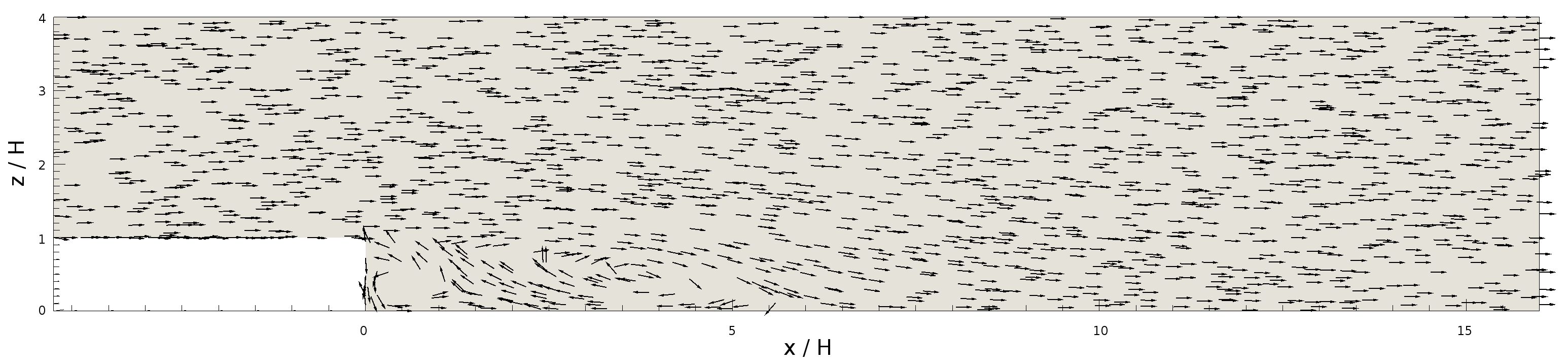}
	\caption{Constant-length arrows indicating time-averaged two-dimensional flow in a vertical slice along the domain. The step is on the bottom left, ending at $x=0$.}
	\label{fig:results-recirculation-arrows}
\end{figure}

Two alternate metrics were used to find the recirculation point: i) the streamwise velocity condition $u=0$, which gave $X_r=6.065 H$, and ii) the skin friction co-efficient conditions $C_f=0$ as per \citet{jovic1994}, which gave a value of $X_r=6.069 H$. These figures are extremely close to each other, with a relative error of 0.067\%. They compare well with the experimental results of \citet{jovic1994}, who reported the reattachment lengths in experiments varying between $X_r=6.0 \pm 0.15 H$, and with \citet{le1993} who found $X_r=6.0 H$. However, this differs from \citet{aider2006}, who found that $X_r$ varied from $5.29 - 5.8 H$; their results were highly dependent upon the treatment of the upstream boundary conditions for turbulence. The DNS simuations of \cite{le1997} gave $X_r=6.28 H$. A summary of all comparisons of $X_r$ can be found in Table \ref{tab:xr_comparison}.
The DG LES values for $X_r$ also match what can be seen in the general features of the time-averaged flow in Figure \ref{fig:results-recirculation-arrows}.  In addition, this figure shows the secondary circulation bubble near the step wall at $x < 1 H$, as noted by \cite{le1997}. 

\begin{table}
\centering
\caption{Comparison of calculated values of $X_r$ with \citet{jovic1994}. Values in brackets () are the smallest possible relative error after accounting for experimental error ($\pm 0.15 H$).}
\begin{tabular}{lll}
\hline\noalign{\smallskip}
Source & $X_r$ (H) & Rel. error (\%)\\
\hline\noalign{\smallskip}
\citet{jovic1994} & 6.0  & -  \\
\citet{le1993} & 6.0 & 0 (0) \\
DG LES ($u=0$) & 6.065 & 1.072 (0)\\
DG LES ($C_f$=0) &  6.069 &  1.137 (0) \\
\citet{aider2006} - WN & 5.8 & 3.448 (0.840) \\
\citet{le1997} & 6.28 & 4.459 (2.121) \\
\citet{aider2006} - PS & 5.29 & 11.833 (9.333) \\
\citet{panjwani2009} & 7.2 & 16.666 (14.583) \\ 
\citet{dubief2000} & 7.2 & 16.666 (14.583) \\ 
\hline\noalign{\smallskip}
\end{tabular}
\label{tab:xr_comparison}
\end{table}

\begin{figure}
\centering
  \includegraphics[width=0.5\linewidth]{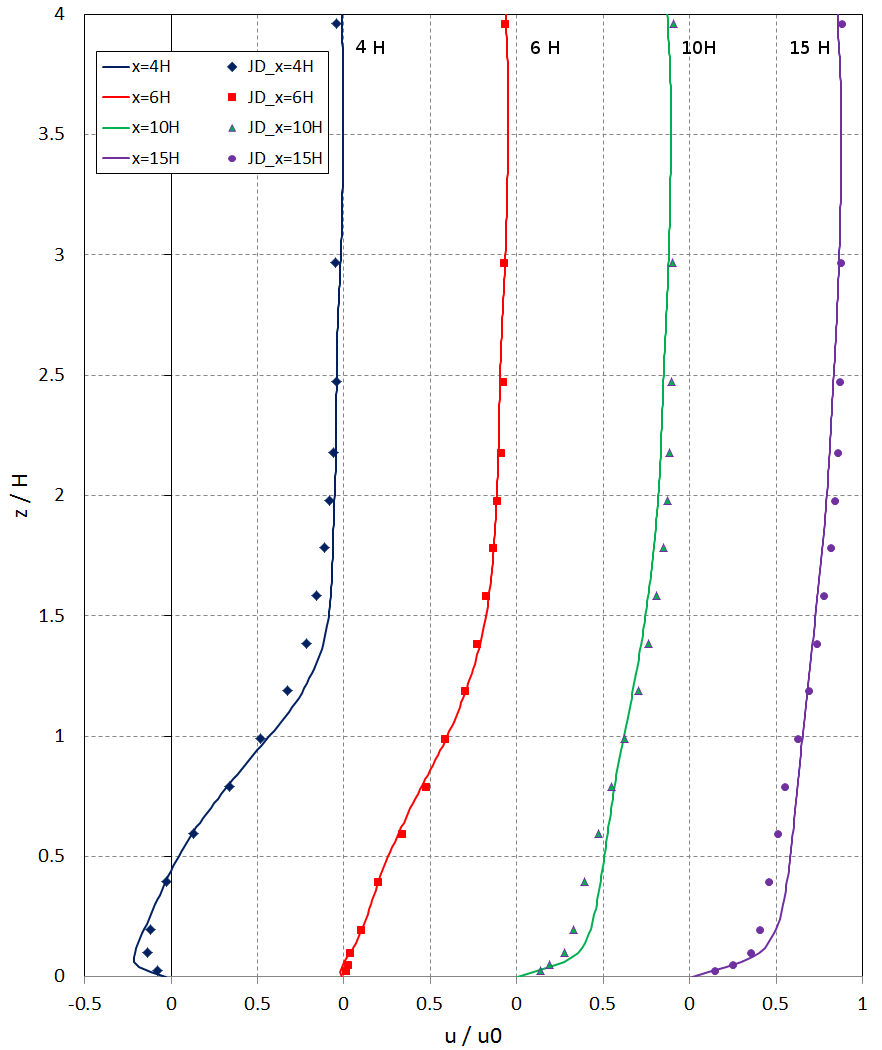}
	\caption{Downstream mean streamwise velocity profiles from the model, versus experimental data \citep{jovic1994}. The solid lines indicate simulation results, and the markers indicate experimental data.}
	\label{fig:vertical-profiles-versus-jovic}
\end{figure}

Vertical velocity profiles of the simulation results can been seen in Figure \ref{fig:vertical-profiles-versus-jovic}, which plots these against the experimental data from \citet{jovic1994}. There is good agreement between them, with the model showing the same characteristic backflow before $X_r$ at $x=4 H$ near the bottom boundary, before recovering to forward flow at $z \approx 0.4 H$. As expected with recirculation zones, there is strong departure from the law of the wall. At $x= 6H$, there is almost no difference between the model and experimental data, whereas at $x=10 H$ and $15 H$, there is a slightly quicker recovery to a logarithmic profile in the model where $z < 0.5 H$. It is possible that this is a shortcoming of the LES model being used: despite the use of Van Driest damping, the value of $C_s$ may be adversely contributing to turbulent diffusion near the wall, as the local value of $C_s$ depends upon the flow regime prevalent there. A dynamic LES scheme \citep{germano1991} may improve this, as $C_s$ will then be calculated and spatially varying, rather than  constant.

\begin{figure}
\centering
  \includegraphics[width=0.99\linewidth]{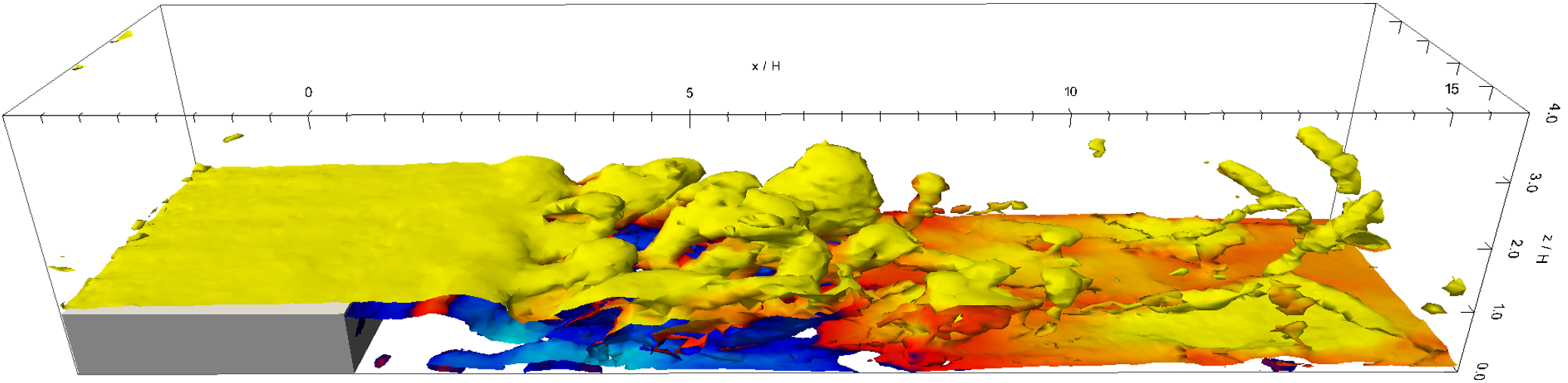}
	\caption{Instantaneous 3D snapshot of vorticity magnitude isosurface (=0.0075 s$^{-1}$), coloured by streamwise flow direction. Yellow indicates forward flow, blue indicates reversed flow.}
	\label{fig:instanteous-vorticity-snapshot}
\end{figure}
\begin{figure}

\centering
  \includegraphics[width=0.95\linewidth]{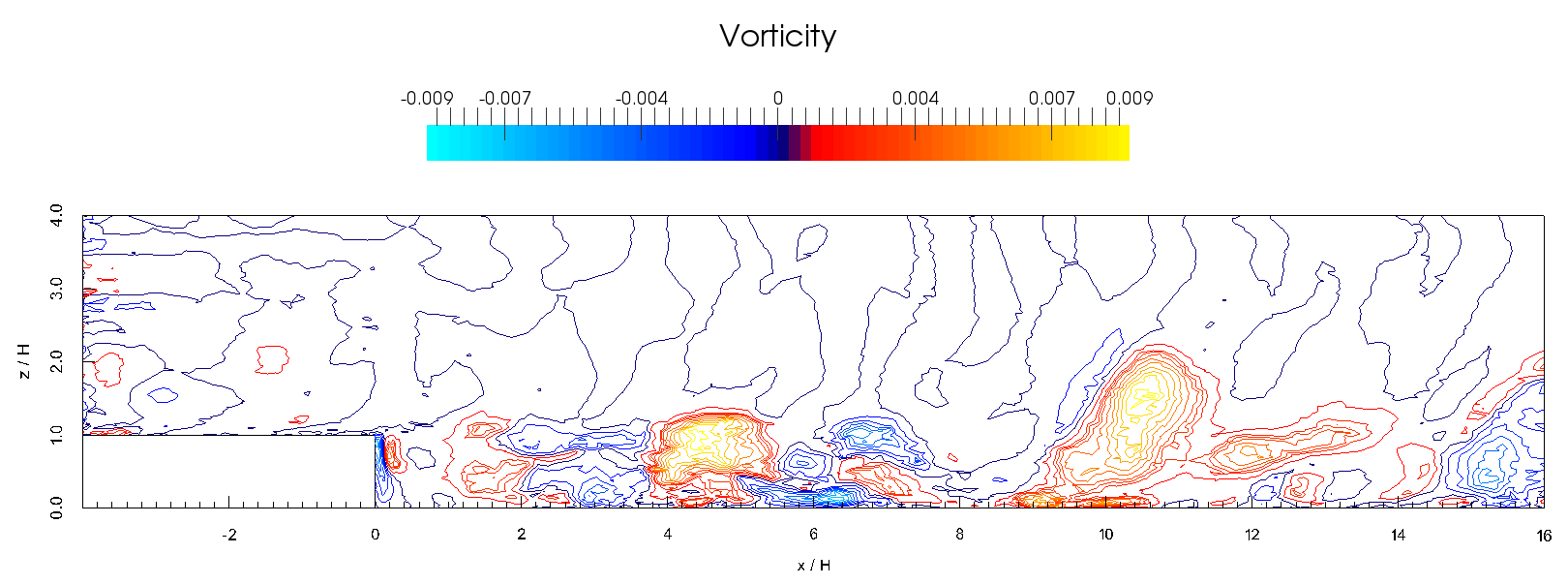}
	\caption{Instanteous vertical slice of the span-wise vorticity contours.}
	\label{fig:vorticity-contours}
\end{figure}

In terms of unsteady flow features, the DG LES model's reattachment point oscillated about the mean $X_r$, as evidenced by the instanteous snapshot in Figure \ref{fig:instanteous-vorticity-snapshot}, which shows the turbulent flow downstream of the step clearly recirculating beyond the predicted average of $6.065 H$. The same behaviour was noted by \citet{friedrich1990} in their LES simulations, and in the experiments of \citet{eaton1980, driver1983, driver1987}, due to the vortical motion downstream of the step. The instanteous snapshot showing spanwise vorticity in Figure \ref{fig:vorticity-contours} demonstrates the same phenomena occuring in the DG LES simulations.

\section{Conclusions}
\label{s:conclusions}

The technique developed here for implementing stable, efficient Large Eddy Simulation for Discontinuous Galerkin finite elements has been presented, with the test case comparing well with experimental data. Whilst the current implementation has been for low order elements, future work will expand this to a general method working with higher orders. It also anticipated that the extension of the work to use a Germano dynamic LES method could further improve already quite acceptable results. Finally, it should be noted that the Discontinuous / Continuous Galerkin mixed element technique used here can not only be applied to LES models, but also to any diagnostic fluid variable dependent on the velocity field, provided that careful attention is paid to the coupling between it and the momentum equation.

\section*{Acknowledgements}

This work used the ARCHER UK National Supercomputing Service. We would like to acknowledge the ARCHER eCSE programme for funding this work, award number eCSE05-7.  Dr. Maddison also acknowledges funding from the UK Natural Environmental Research Council (NE/L005166/1).

\bibliography{tidal}
\bibliographystyle{plainnat}

\end{document}